\documentclass[english,letterpaper,aps,prl,twocolumn,showpacs]{revtex4}
\usepackage{lmodern}

\usepackage[T1]{fontenc}
\usepackage[latin1]{inputenc}
\usepackage{graphicx}
\usepackage{amssymb}

\makeatletter

\usepackage{lmodern}

\makeatletter

\usepackage{lmodern}

\makeatletter

\usepackage{lmodern}

\makeatletter

\usepackage{lmodern}

\makeatletter

\makeatletter

\usepackage{epsfig}

\usepackage{dcolumn}

\usepackage{bm}

\usepackage{bbm}

\usepackage{wasysym}

\usepackage{marvosym}

\newlength{\textwidthm}

\makeatother

\makeatother

\makeatother

\makeatother

\makeatother

\usepackage{babel}
\makeatother

\begin{document}

\title{Kondo Quantum Criticality of Magnetic Adatoms in Graphene}

\author{Bruno Uchoa$^{1}$, T. G. Rappoport$^{2}$, and A.~H. Castro Neto$^{3}$}

\affiliation{Department of Physics$\mbox{,}$ University of Illinois at Urbana-Champaign$\mbox{,}\,$1110
W. Green St, Urbana, IL, 61801, USA}

\affiliation{$^{2}$Instituto de F\'{\i}sica, Universidade Federal do Rio de Janeiro,
Rio de Janeiro, RJ, 68.528-970, Brazil}

\affiliation{$^{3}$Department of Physics, Boston University, 590 Commonwealth
Avenue, Boston, MA 02215, USA}

\date{\today}

\begin{abstract}
We examine the exchange Hamiltonian for magnetic adatoms in graphene
with localized inner shell states. On symmetry grounds, we predict
the existence of a class of orbitals that lead to a distinct class
of quantum critical points in graphene, where the Kondo temperature
scales as $T_{K}\propto|J-J_{c}|^{1/3}$ near the critical coupling
$J_{c}$ , and the local spin is effectively screened by a \emph{super-ohmic}
bath. For this class, the RKKY interaction decays spatially with a
fast power law $\sim1/R^{7}$. Away from half filling, we show that
the exchange coupling in graphene can be controlled across the quantum
critical region by gating. We propose that the vicinity of the Kondo
quantum critical point can be directly accessed with scanning tunneling
probes and gating. 
\end{abstract}


\maketitle
Graphene is a single atomic sheet of carbon atoms with elementary
electronic quasiparticles that behave as massless Dirac fermions\cite{Novo}.
The Kondo effect has been recently observed in graphene\cite{manoharan,Chen},
and the formation of a Kondo screening cloud around a magnetic adatom
is quantum critical at half filling\cite{fradkin90,Zhang}, crossing
over at weak coupling to the standard Fermi liquid case, when the
DOS is locally restored by disorder\cite{Hentschel} or gating effects\cite{Sengupta}.
The Kondo resonance in graphene is also strongly sensitive to the
position of the adatom in the honeycomb lattice, where the interplay
of orbital and spin degrees of freedom may give rise to an SU(4) Kondo
effect\cite{Wehling2}.

In this letter, after establishing a generic one-level exchange interaction
Hamiltonian for magnetic adatoms in graphene, we show there is a symmetry
class of orbitals in which quantum interference between the different
hybridization paths leads to a fixed point where the Kondo temperature
$T_{K}\propto|J-J_{c}|^{\nu}$, scales with the mean field exponent
$\nu=1/3$, with $J$ as the Kondo coupling near criticality. In the
$\nu=1/3$ class, graphene behaves as a \emph{super-ohmic} bath for
the local spin and the RKKY interaction is strongly suppressed, decaying
spatially with a fast power law $\sim1/R^{7}$. Furthermore, we show
that the exchange coupling in graphene can be controlled by gating.
This effect opens the possibility of exploring the proximity to the
Kondo quantum critical point (QCP) in graphene directly with scanning
tunneling probe (STM) measurements\cite{Uchoa09,Zhuang,sengupta2,Wehling}.

We start from the graphene Hamiltonian, $\mathcal{H}_{g}=-t\sum_{\sigma,\langle ij\rangle}a_{\sigma}^{\dagger}(\mathbf{R}_{i})b_{\sigma}(\mathbf{R}_{j})+{\rm h.c.},$
where $a,b$ are fermionic operators on sublattices $A$ and $B$,
respectively, $t\sim2.8$ eV is the nearest neighbors hopping energy
and $\sigma=\uparrow\downarrow$ labels the spin. In the momentum
space,\begin{equation}
\mathcal{H}_{g}=-t\sum_{\mathbf{p}\sigma}\phi_{\mathbf{p}}a_{\sigma,\mathbf{p}}^{\dagger}b_{\sigma,\mathbf{p}}+h.c.,\label{eq:Hg}\end{equation}
 where $\phi_{\mathbf{p}}=\sum_{i=1}^{3}\mbox{e}^{i\mathbf{p}\cdot\mathbf{a}_{i}}$,
and $\mathbf{a}_{1}=\hat{x}$, $\mathbf{a}_{2}=-\hat{x}/2+\sqrt{3}\hat{y}/2$,
and $\mathbf{a}_{3}=-\hat{x}/2-\sqrt{3}\hat{y}/2$ are the lattice
nearest neighbor vectors.

In the presence of a localized level, the problem is described by
the single impurity Anderson Hamiltonian\cite{anderson61,Uchoa08},
$\mathcal{H}=\mathcal{H}_{g}+\mathcal{H}_{f}+\mathcal{H}_{U}+\mathcal{H}_{V}$,
where $\mathcal{H}_{f}=\sum_{\sigma}\epsilon_{0}\,\hat{n}_{f,\sigma}$
is the Hamiltonian of the localized electrons, with $\hat{n}_{f,\sigma}=f_{\sigma}^{\dagger}f_{\sigma}$
as the number operator, and $\epsilon_{0}$ is the energy of the local
state measured relative to the Dirac point, $\mathcal{H}_{U}=U\hat{n}_{f,\uparrow}\hat{n}_{f,\downarrow}$
gives the electronic repulsion in the localized level, and $H_{V}$
describes the hybridization between the local level and the graphene
electrons.

\begin{figure}[b]
\begin{centering}
\includegraphics[scale=0.21]{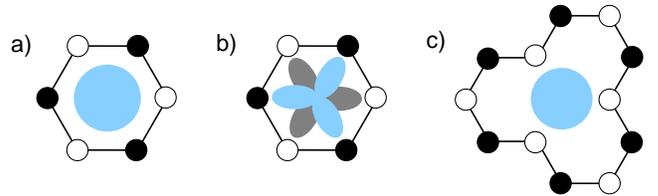} 
\par\end{centering}

\caption{{\small Representation of (a) an $s$-wave and (b) an $f$-wave orbital,
when the adatom sits in the center of the graphene honeycomb hexagon.
(c) Substitutional impurity in a single vacancy. In the three cases,
the adatom hybridizes equally with the neighboring carbons on the
same sublattice. }}

\end{figure}

The adatoms in graphene can sit for instance on top of a carbon atom,
where the hybridization Hamiltonian is $\mathcal{H}_{V}^{A}\!=\! V\sum_{\sigma} a_{\sigma}^{\dagger}(0)f_{\sigma}+ h.c.$,
or in the hollow site in the center of the honeycomb hexagon, where
the adatom hybridizes with the two sublattices, $\mathcal{H}_{V}^{H}\!=\!\sum_{\sigma}\sum_{i=1}^{3}\left[V_{a,i}a_{\sigma}^{\dagger}(\mathbf{a}_{i})+V_{b,i}b_{\sigma}^{\dagger}(-\mathbf{a}_{i})\right]f_{\sigma}+h.c.$,
with $V_{x,i}$ ($x=a,b$) representing the hybridization strength
of the localized orbital with each of the three surrounding carbon
atoms sitting on a given sublattice. In momentum representation \cite{Uchoa09},\begin{equation}
\mathcal{H}_{V}=\sum_{\mathbf{p}\sigma}\left[V_{b,\mathbf{p}}b_{\sigma,\mathbf{p}}^{\dagger}+V_{a,\mathbf{p}}^{*}a_{\sigma,\mathbf{p}}^{\dagger}\right]f_{\sigma}+ h.c.\,,\label{eq:baf}\end{equation}
 where $V_{a\mathbf{p}}\equiv V\,,$ and $V_{b,\mathbf{p}}=0$ for
a top carbon site, say on sublattice $A$ ($A$-site). When the adatom
sits in the center of the hexagon ($H$-site), or else for a substitutional
impurity in a single vacancy\cite{kras} ($S$-site), the hybridization
function is $V_{x,\mathbf{p}}=\sum_{i=1}^{3}V_{x,i}\mbox{e}^{i\mathbf{p}\cdot\mathbf{a}_{i}}.$
On $H$-sites, for an $s$-wave orbital, $V_{x,i}\equiv V$, giving
$V_{x,\mathbf{p}}\equiv V\phi_{\mathbf{p}}$, whereas for an in-plane
$f$-wave orbital, as shown in Fig. 1b, where the orbital is odd in
the two sublattices, $V_{b,i}=-V_{a,i}\equiv V$, resulting in $-V_{a,\mathbf{p}}=V_{b,\mathbf{p}}\equiv V\phi_{\mathbf{p}}$.
In the case of an $s$ or in-plane $f$-wave orbital on an $S$-site
on sublattice $A$, $V_{a,\mathbf{p}}=0$, and $V_{b,\mathbf{p}}\equiv V\phi_{\mathbf{p}}$,
whereas for a substitutional impurity on a $B$-site, $V_{a,\mathbf{p}}\equiv V\phi_{\mathbf{p}}$,
and $V_{b,\mathbf{p}}=0$ (see Fig. 1c).

Diagonalizing the non-interacting part of the Hamiltonian $\mathcal{H}$
in the $A$, $B$ sublattices, \begin{equation}
\mathcal{H}_{g}=\sum_{\mathbf{p}\alpha\sigma}\, E_{\alpha}(\mathbf{p})c_{\alpha,\sigma,\mathbf{p}}^{\dagger}c_{\alpha,\sigma,\mathbf{p}}\,,\label{eq:Hg2}\end{equation}
 where $E_{\alpha}(\mathbf{p})=\alpha t|\phi_{\mathbf{p}}|$ is the
graphene tight-binding spectrum, with $\alpha=\pm$ labeling the conduction
and valence bands, and $c_{\pm,\sigma,\mathbf{k}}=(b_{\sigma,\mathbf{k}}\pm\phi_{\mathbf{k}}^{*}/|\phi_{\mathbf{k}}|a_{\sigma,\mathbf{k}})/\sqrt{2}$
are the new operators in the diagonal basis. The hybridization term
in the rotated basis is\begin{equation}
\mathcal{H}_{V}=V\sum_{\alpha=\pm}\sum_{\mathbf{p},\sigma}\left[\Theta_{\alpha,\mathbf{p}}c_{\alpha,\mathbf{p}\sigma}^{\dagger}f_{\sigma}+h.c.\right],\label{eq:HV}\end{equation}
 where \begin{equation}
\Theta_{\alpha,\mathbf{p}}=(V_{b,\mathbf{p}}+\alpha V_{a,\mathbf{p}}^{*}\phi_{\mathbf{p}}^{*}/|\phi_{\mathbf{p}}|)/(\sqrt{2}V).\label{eq:Theta}\end{equation}
 In particular, $\Theta_{\alpha,\mathbf{p}}^{A}=1/\sqrt{2}$ when
the adatom is on top of an $A$-site, $\Theta_{\alpha,\mathbf{p}}^{B}=\alpha\phi_{\mathbf{p}}^{*}/(\sqrt{2}|\phi_{\mathbf{p}}|)$
on a $B$-site, and $\Theta_{\alpha,\mathbf{p}}^{H,\gamma}=[\phi_{\mathbf{p}}+(-1)^{\gamma}\alpha\phi_{\mathbf{p}}^{*2}/|\phi_{\mathbf{p}}|]/\sqrt{2}$
when the adatom sits on an $H$-site, where $\gamma=0$ for an $s$-wave
orbital and $\gamma=1$ in-plane $f$-wave orbital. In the substitutional
case, $\Theta_{\alpha,\mathbf{p}}^{S_{A}}=\phi_{\mathbf{p}}/\sqrt{2}$
for an impurity on sublattice $A$, and $\Theta_{\alpha,\mathbf{p}}^{S_{B}}=\alpha\phi_{\mathbf{p}}^{*2}/(|\phi_{\mathbf{p}}|\sqrt{2})$
on sublattice $B$.

For all possible symmetries, the orbitals of adatoms sitting on $S$
or $H$ sites can be classified among those that either break or preserve
the $C_{3v}$ point group symmetry of the triangular sublattice in
graphene. Since $|\phi_{\mathbf{p}}|$ scales with $|\omega|/t$,
the orbital level broadening, $\Delta(\omega)=\pi V^{2}\sum_{\alpha,\mathbf{p}}|\Theta_{\alpha,\mathbf{p}}|^{2}\delta(\omega-\alpha t|\phi_{\mathbf{p}}|)$
is either $\Delta(\omega)\propto V^{2}\rho(\omega)$ for orbitals
that explicitly break the $C_{3v}$ point group symmetry, in which
case $|\Theta_{\alpha,\mathbf{p}}|$ scales to a constant near the
Dirac points, where $\rho\propto|\omega|$ is the graphene density
of states (DOS), or else $\Delta(\omega)\propto V^{2}\rho(\omega)|\omega|^{2}/t^{2}$,
for $C_{3v}$ invariant orbitals, when $|\Theta_{\alpha,\mathbf{p}}|\propto|\phi_{\mathbf{p}}|$
scales to zero at small energy. The first class of orbitals, where
$\Delta(\omega)\propto|\omega|$ (say, type I), represents the standard
case of \textit{ohmic} dissipation\cite{leggett}, and is described
for instance by adatoms on top carbon sites, by $E_{1}$($d_{xz},$$d_{yz}$)
and $E_{2}$$(d_{xy},d_{x^{2}-y^{2}})$ representations of $d$-wave
orbitals and $f_{xz^{2}}$, $f_{yz^{2}}$, $f_{xyz}$, $f_{z(x^{2}-y^{2})}$
orbitals in $H$/$S$ sites. The second class, where $\Delta(\omega)\propto|\omega|^{3}/t^{2}$
(type II), represents a new class of \textit{super-ohmic} dissipation\cite{leggett},
and is described by $s$, $d_{zz}$, $f_{z^{3}}$, $f_{x(x^{2}-3y^{2})}$,
and $f_{y(3x^{2}-y^{2})}$ orbitals in $H$ or $S$ sites (see Fig.1),
where the adatom hybridizes equally with the three nearest carbon
atoms on a given sublattice. On physical grounds, this new class emerges
from quantum mechanical interference between the different hybridization
paths in the honeycomb lattice, as the electrons hop in and out of
the localized level. As we will show, these two classes of orbitals
are described by two distinct types of Kondo QCP.

The Anderson Hamiltonian in graphene can be separated in two terms,
$\mathcal{H}=\mathcal{H}_{0}+\mathcal{H}_{V}$, and then mapped into
a spin exchange Hamiltonian through a standard canonical transformation,
$\bar{\mathcal{H}}=\mbox{e}^{S}\mathcal{H}\mbox{e}^{-S}=\mathcal{H}+[S,\mathcal{H}]+\frac{1}{2}[S,[S,\mathcal{H}]]+...\,$,
where $S=-\sum_{\mathbf{p},\alpha\sigma}V[(1-\hat{n}_{f,-\sigma})(\epsilon_{0}-\alpha t|\phi_{\mathbf{p}}|)^{-1}+\hat{n}_{f,-\sigma}(\epsilon_{0}+U-\alpha t|\phi_{\mathbf{p}}|)^{-1}]\Theta_{\alpha,\mathbf{p}}c_{\alpha,\mathbf{p}\sigma}^{\dagger}f_{\sigma}-h.c.\,,$
which results in a Hamiltonian that is quadratic in $V$ to leading
order, $\bar{\mathcal{H}}=\mathcal{H}_{0}+[S,\mathcal{H}_{V}]/2+O(V^{3})$\cite{Schrieffer}.
At large $U$, the exchange Hamiltonian is given by \begin{equation}
\mathcal{H}_{e}=-J\sum_{\mathbf{k}\mathbf{k}^{\prime}}\sum_{\alpha\alpha^{\prime}}\Theta_{\alpha\mathbf{k}}^{*}\Theta_{\alpha^{\prime}\mathbf{k}^{\prime}}\,\mathbf{S}\cdot c_{\alpha^{\prime},\sigma^{\prime},\mathbf{k}^{\prime}}^{\dagger}\vec{\sigma}c_{\alpha,\sigma,\mathbf{k}}\,,\label{eq:He}\end{equation}
 where $\vec{\sigma}=(\sigma_{1},\sigma_{2},\sigma_{3})$ are Pauli
matrices and \begin{equation}
J(\mu)\approx V^{2}U/[(\epsilon_{0}-\mu)(\epsilon_{0}+U-\mu)]<0\,,\label{eq:J}\end{equation}
 is the exchange coupling defined at the Fermi level, $\mu$.

The validity of the exchange Hamiltonian (\ref{eq:He}) is controlled
by the ratio $\Delta(\epsilon_{0})/|\epsilon_{0}-\mu|\ll1$, when
the valence of the localized level is unitary (and hence, the local
spin is a good quantum number) and perturbation theory is well defined
in the original Anderson parameters\cite{note3}. In graphene, where
$\Delta(\omega)\propto\pi V^{2}\rho(\omega)(|\omega|/t)^{\eta}$,
($\eta=0,$ or $2$), with $\rho(\omega)=|\omega|/D^{2}$, and $D\sim7$eV
as the bandwidth, this criterion becomes $|J|\sim V^{2}/(\mu-\epsilon_{0})\ll D^{2}t^{\eta}/(\pi|\epsilon_{0}|^{1+\eta})$.
When the level is exactly at the Dirac point, $\epsilon_{0}=0$, the
level broadening is zero\cite{Uchoa08} and the exchange coupling
$|J|\sim V^{2}/\mu$ has no upper bound and can be shifted by gating
towards the strong coupling limit of the Kondo problem, $J\to\infty$,
when the Fermi level is tuned to the Dirac point, $\mu\to0^{+}$\cite{note0}.
Since the experimentally accessible range of gate voltage for graphene
on a 300 nm thick SiO$_{2}$ substrate is $\mu\in[-0.3,0.3]$ eV,
the exchange coupling of a magnetic Co adatom, for instance, with
$V=1$ eV and $\epsilon_{0}=-0.4$ eV, can be tuned continuously in
the range between $|J|\in1.4-10$ eV. This effect, which is allowed
by the low DOS in graphene, brings the unprecedented experimental
possibility of controlling the exchange coupling and switching magnetic
adatoms between different Kondo coupling regimes in the proximity
of a QCP, as we show in Fig. 2a.

Since the determinant of the exchange coupling matrix in Eq. (\ref{eq:He}),
$\mbox{det}[\hat{J}_{\alpha\alpha^{\prime}}]$, is identically zero,
the exchange Hamiltonian (\ref{eq:He}) can be diagonalized into a
new basis where one of the channels decouples from the bath\cite{Pulstilnik}.
The eigenvalues in the new basis are $J_{u,\mathbf{k},\mathbf{k}^{\prime}}=J\sum_{\alpha}\Theta_{\alpha\mathbf{k}}^{*}\Theta_{\alpha\mathbf{k}^{\prime}}$
and $J_{v}=0$, and hence, the generic one-level exchange Hamiltonian
(\ref{eq:He}) maps into the problem of a \emph{single} channel Kondo
Hamiltonian, $\mathcal{H}_{e}=-2\sum_{\mathbf{k},\mathbf{k}^\prime }J_{u,\mathbf{k}\mathbf{k}^{\prime}}\mathbf{S}\cdot\mathbf{s}_{\mathbf{k},\mathbf{k}^{\prime}}$,
where $\mathbf{s}$ is the itinerant spin, regardless the implicit
valley degeneracy.

In the one-level problem, the renormalization of the constant $J$
due to the coupling of the local spin with the bath is given by: $J^{\prime}=J-2N_{s}J^{2}\rho(D)(D/t)^{\eta}\,\delta D/D$,
after integrating out the high energy modes with energy $D$ at the
bottom of the band, where $N_{s}=1,\,2$ describes the number of sublattices
the adatom effectively hybridizes. Since a DOS in the form $\rho(\omega)\propto|\omega|^{r}$
has a scaling dimension $r$, where $r=1$ in graphene, the restoration
of the cut-off in the {}``poor man's scaling'' analysis requires
an additional rescaling $J^{\prime}\to[(D+\delta D)/D]^{r+\eta}J^{\prime}$\cite{fradkin90},
which results in the beta function\begin{equation}
\beta(J)=\frac{\mbox{d}J}{\mbox{d}\ln D}=-(r+\eta)J-2N_{s}J^{2}\rho(D)(D/t)^{\eta}\,.\label{eq:beta}\end{equation}
 The renormalization group (RG) flow leads to a intermediate coupling
(IC) fixed point at $J_{c}=-(r+\eta)t^{\eta}/[2N_{s}\rho(D)D^{\eta}]$,
which separates the weak and strong coupling sectors. For type I orbitals
(ohmic bath), one recovers the usual IC fixed point $J_{c}=-r/[2N_{s}\rho(D)]$\cite{fradkin90},
whereas for type II (super-ohmic bath, $\eta=2$) $J_{c}\approx-3t^{2}/(2N_{s}D)$
in the Dirac case ($r=1$). In graphene, this new fixed point describes
a one-channel Kondo problem in the presence of an effective fermionic
bath with DOS $\rho\propto|\omega|^{\bar{r}}$, where $\bar{r}\equiv r+\eta=3$.
Since the tree level scaling dimension of the hybridization $V$ in
the Anderson model is $(1-\bar{r})/2$, the case $\bar{r}=1$ corresponds
to an upper critical scaling dimension, above which ($\bar{r}>1$)
$V$ is an irrelevant perturbation in the RG sense\cite{Vojta2}.
In this situation, fluctuations are not important near the QCP, and
the critical exponents are expected to be \emph{mean-field like},
in contrast with the marginal case ($\bar{r}=1$), where mean field
cannot be trusted\cite{Vojta}.

The RG analysis derived from the exchange Hamiltonian (\ref{eq:He})
can be verified directly from the hybridization Hamiltonian (\ref{eq:baf}).
In the large $N$ limit near the critical regime, singly occupied
level states are enforced at the mean field level through the constraint
$\lambda(\sum_{m}^{N}f_{m}^{\dagger}f_{m}-1)=0$\cite{Newns}, with
$N=2$ in the spin 1/2 case. The minimization of the energy $\partial\langle\mathcal{H}\rangle/\partial\lambda=0$
gives $\lambda=\frac{N}{\pi}\int_{-\infty}^{\infty}\mbox{d}\omega\, n(\omega)\,\mbox{Im}[G_{ff}(\omega)\Sigma_{ff}(\omega)],$
where $n(\omega)=[\mbox{e}^{(\omega-\mu)/T}+1]^{-1}$ is the Fermi
distribution, $T$ is the temperature, and $G_{ff}(\tau)=-\langle T[f_{\sigma}(\tau)f_{\sigma}^{\dagger}(0)]\rangle$
is the $f$-electrons Green's function, $G_{ff}(\omega)=[\omega-\epsilon_{0}-\lambda+\Sigma_{ff}(\omega)+i0^{+}]^{-1}$.
$\Sigma_{ff}(\omega)=V^{2}\sum_{\alpha,\mathbf{p}}|\Theta_{\alpha,\mathbf{p}}|^{2}\,\hat{G}_{\alpha\mathbf{p}}^{0\, R}(\omega)$
is the self-energy of the $f$-electrons in the presence of the graphene
bath, where $\Delta(\omega)=-\mbox{Im}\Sigma_{ff}(\omega)$ defines
the level broadening and $G_{\alpha,\mathbf{p}}^{0\, R}(\omega)=(\omega-\alpha|\phi_{\mathbf{p}}|+i0^{+})^{-1}$
is the retarded Green's function of the $c$-electrons in the diagonal
basis, $G_{\alpha}^{0}(\tau)=-\langle T[c_{\alpha}(\tau)c_{\alpha}^{\dagger}(0)]\rangle$.
In the critical regime, where $\lambda=\mu-\epsilon_{0}\equiv V^{2}/|J|$,
the Kondo temperature in graphene is extracted to leading order in
$V$ from the equation \begin{equation}
\frac{1}{J}=-\frac{N}{2}\sum_{\mathbf{p},\alpha}\frac{|\Theta_{\alpha,\mathbf{p}}|^{2}}{\alpha t|\phi_{\mathbf{p}}|+\mu}\tanh\!\left[\frac{\alpha t|\phi_{\mathbf{p}}|+\mu}{2T_{K}}\right].\label{eq:JEq}\end{equation}

In the Dirac cone approximation, the Kondo temperature for orbitals
of type II ($\eta=2$) is \begin{equation}
T_{K}=(D/2)\left(1-J_{c}/J\right)^{1/3}\label{eq:1/3}\end{equation}
 at half filling, where $J_{c}=-3t^{2}/(N_{s}ND)$ is the same critical
coupling derived from the RG equation (\ref{eq:beta}). Away from
half filling, $J_{c}$ defines the crossover between the Fermi liquid
weak coupling regime, at $J/J_{c}\ll1$, where $T_{K}=(|\mu|/2)\exp[D^{3}/(3|\mu|^{3})(1-J_{c}/J(\mu)+3\mu^{2}/D^{2})]$,
and the strong coupling regime, for $|J|\gtrsim|J_{c}|\approx(2/N_{s})$eV,
where $T_{K}\approx(D/2)[1-J_{c}/J(\mu)+3\mu^{2}/D^{2}]^{1/3}$, as
shown in Fig. 2b. At the critical coupling $(J=J_{c})$, \begin{equation}
T_{K}=(D/2)|3\mu^{2}/D^{2}|^{1/3}\,,\label{eq:Tkmu}\end{equation}
 and the fingerprint of the QCP at $\mu=0$ can be observed in the
scaling of the Kondo temperature with $\mu$ in the vicinity of the
QCP, at $J\sim J_{c}$. This scaling can be measured in STM, where
the signature of the Kondo effect is manifested in the form of a Kondo
resonance in the DOS at the Fermi level, for $T<T_{K}$. %
\begin{figure}
\begin{centering}
\includegraphics[scale=0.37]{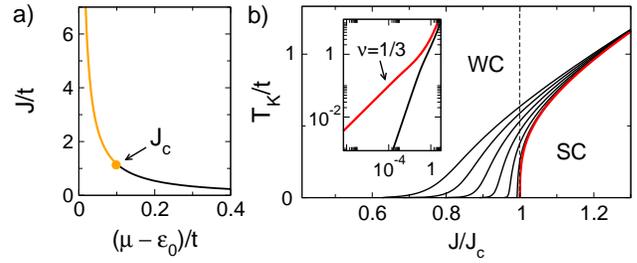} 
\par\end{centering}

\caption{{\small (color online) a) Kondo coupling vs. gate $\mu$ for $U/t=1$
and $V/t=0.3$. The dot illustrates a typical value for the critical
coupling $J_{c}$. b) Kondo temperature $T_{K}/t$ vs. $J$ for $C_{3v}$
invariant orbitals. Red (light) curve: $\mu=0$; black: $\mu/D=$0.05,
0.1, 0.15, 0.2, $0.25$, and 0.3. The line $J\sim J_{c}$ sets the
crossover scale between the Kondo weak coupling (WC) and strong coupling
(SC) regimes. Inset: $T_{K}/t$ vs. $J/J_{c}-1$ near the QCP ($\mu=0$),
in log scale. Red (light) line: type II orbitals ($\nu=1/3$); black:
type I ($\nu=1$) (see text).}}

\end{figure}

In Fig. 2b, we numerically calculate the scaling of the Kondo temperature
in tight-binding. For type II orbitals, the $\nu=1/3$ exponent in
the Kondo temperature, $T_{K}\propto|J-J_{c}|^{\nu}$, found in the
linear cone approximation persists above room temperature, up to $T_K/t\sim 1$ (red curves). 
In the more standard ohmic case, for spins on top carbon
sites (black curve of the inset), the scaling is linear ($\nu=1$)
at the mean-field level.

Tracing the conduction electrons in the exchange Hamiltonian (\ref{eq:He}),
the RKKY Hamiltonian of a spin lattice in graphene is $H_{RKKY}=-J^{2}\sum_{ij}\chi_{ij}^{x,y}\,\mathbf{S}_{i}\cdot\mathbf{S}_{j}\,,$
where $\chi_{ij}^{x,y}\equiv\chi^{x,y}(\mathbf{R}_{i}-\mathbf{R}_{j})$
is the spin susceptibility, with $i,j$ indexing the local spins,
and $x,y=A,\, B,\, H,\, S_{A},\, S_{B}$ label the position of the
magnetic adatoms in lattice. In momentum space,\begin{equation}
\chi^{x,y}(\mathbf{q})=\sum_{\alpha\alpha^{\prime},\mathbf{p}}\mathcal{M}_{\alpha,\alpha^{\prime}\mathbf{p},\mathbf{q}}^{x,y}\frac{n[E_{\alpha^{\prime}}(\mathbf{p}+\mathbf{q})]-n[E_{\alpha}(\mathbf{p})]}{E_{\alpha^{\prime}}(\mathbf{p}+\mathbf{q})-E_{\alpha}(\mathbf{p})}\,,\label{eq:chi3}\end{equation}
 where $\mathcal{M}_{\alpha,\alpha^{\prime},\mathbf{p},\mathbf{q}}^{xy}=\Theta_{\alpha\mathbf{p}}^{*x}\Theta_{\alpha\mathbf{p}}^{y}\Theta_{\alpha^{\prime}\mathbf{p}+\mathbf{q}}^{x}\Theta_{\alpha^{\prime}\mathbf{p}+\mathbf{q}}^{*y}$.

\begin{figure}[t]

\begin{centering}
\includegraphics[scale=0.34]{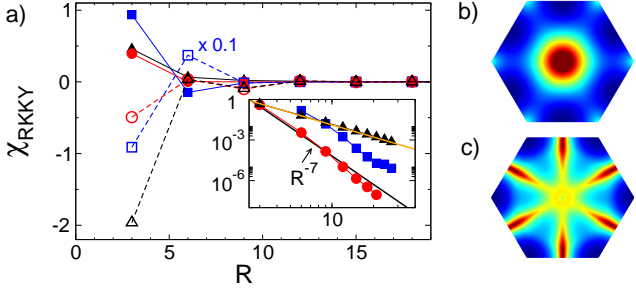} 
\par\end{centering}

\caption{{\small (color online) a) $\chi(R)$ vs distance $R$, along a zigzag
direction (in lattice units), for $A$-sites (black triangles), $H$-sites
(blue squares), and $S$ sites for spins on the same sublattice (red
circles). Solid lines: $\mu=0$; dashed: $\mu=t$. Inset: $|\chi_{ij}|$
plot in a log scale. Orange (light) guide line: $1/R^{3}$; black:
$1/R^{7}$. On the right: $\chi(\mathbf{q})$ for $A$ site spins,
plotted in the graphene BZ at b) $\mu=0$ and c) $\mu=t$. Red (dark)
regions represent $\chi(\mathbf{q})>0$ and blue (light) regions $\chi(\mathbf{q})<0$.}}

\end{figure}

For spins on the same sublattice, $\mathcal{M}^{AA}\!=\!1/4$, whereas
on opposite sublattices $\mathcal{M}^{AB}\!=\!\alpha\alpha^{\prime}\phi_{\mathbf{p}}\phi_{\mathbf{p}+\mathbf{q}}^{*}/(4|\phi_{\mathbf{p}}||\phi_{\mathbf{p}+\mathbf{q}}|)$,
in agreement with Ref.\cite{Brey}, in the Dirac cone limit. For
an $H$-site\cite{note}, \begin{equation}
\mathcal{M}^{HH}=|\Theta_{\alpha,\mathbf{p}}^{H}|^{2}|\Theta_{\alpha^{\prime},\mathbf{p}+\mathbf{q}}^{H}|^{2},\label{eq:Mhh}\end{equation}
 where $|\Theta_{\alpha,\mathbf{p}}^{H,\gamma}|^{2}=|\phi_{\mathbf{p}}|^{2}\left[1+(-1)^{\gamma}\alpha\mbox{Re}(\phi_{\mathbf{p}}^{3})/|\phi_{\mathbf{p}}|^{3}\right]$
for orbitals of type II; for $S$-sites, $\mathcal{M}^{S_{A}S_{A}}=|\phi_{\mathbf{p}}|^{2}|\phi_{\mathbf{p}+\mathbf{q}}|^{2}/4$
for spins on the same sublattice, and $\mathcal{M}^{S_{A}S_{B}}=\alpha\alpha^{\prime}\phi_{\mathbf{p}}^{3}\phi_{\mathbf{p}+\mathbf{q}}^{*\,3}/(4|\phi_{\mathbf{p}}||\phi_{\mathbf{p}+\mathbf{q}}|)$
for opposite ones.

In Fig. 3a, we show the spacial decay of the RKKY interaction on the
lattice for type II orbitals on $A$, $H$ and $S_{A}$ site spins.
At half filling, the RKKY interaction is always ferromagnetic for
same sublattice spins, substitutional or not, and antiferro for spins
on opposite sublattices\cite{Saremi,Brey}. The $H$ case on the
other hand, is ferromagnetic for nearest neighbor spins and antiferromagetic
at longer distances (blue squares). In the $H$ and $S$ cases, the
interaction is short ranged and decays with a fast power law $\sim1/R^{7}$,
in contrast to the known $1/R^{3}$ decay in the $A$ site case\cite{Vozmediano,Cheianov,Saremi,Brey},
as shown in the inset of Fig. 3. This fast decay is consistent with
the case of carbon nanotubes, where the RKKY interaction decays with
$1/R$ for top carbon sites and with $1/R^{5}$ for isotropic orbitals
on $H$ sites\cite{Kirwan}.

Fig. 3b and 3c display the magnetic peaks in the susceptibility in
the $A$-site case for $\mu=0$, and $\mu=t$. For $\mu<t$, $\chi(\mathbf{q})$
has a strong ferromagnetic forward scattering peak around the center
of the BZ ($q=0$), and six subdominant antiferromagnetic peaks at
corners of the BZ. Exactly at $\mu=t$, a strong peak emerges at the
$M$ point due to the nesting of the Van-Hove singularities (VHS)
of the graphene band (see Fig. 3c), where the DOS diverges logarithmically.
This peak reverses the ordering pattern of the RKKY interaction in
comparison to the $\mu=0$ regime in all studied cases, as shown in
the dashed lines of Fig. 3a. When $\mu$ is at the VHS, the interaction
between spins on same (opposite) sublattices, substitutional or not,
is always antiferromagnetic (ferro). In the same way, the RKKY interaction
in the $H$ site case becomes antiferromagnetic for nearest neighbor
sites and ferromagnetic at long distances.

In conclusion, we have derived the one-level exchange Hamiltonian
for magnetic adatoms in graphene and shown the existence of two symmetry
classes of magnetic orbitals that correspond to distinct classes of
Kondo QCP. We also showed that the exchange coupling can be controlled
across the quantum critical region with the application of a gate
voltage.

We thank E. Fradkin, A. Balatsky, L. Brey and S. Lal for discussions.
BU acknowledges partial support from the DOI grant DE-FG02-91ER45439
at University of Illinois. TGR acknowledges support of CNPq and FAPERJ.
AHCN acknowledges DOE grant DE-FG02-08ER46512.

\newpage

\begin{widetext}

\begin{center}
\textbf{\large Erratum: Kondo Quantum Criticality of Magnetic Adatoms
in Graphene }\\
\textbf{\large {[}Phys. Rev. Lett. 106, 016801 (2011)]}
\par\end{center}{\large \par}

\begin{center}
Bruno Uchoa, T. G. Rappoport, and A.~H. Castro Neto
\end{center}

In Ref.\cite{uchoa_E} we have predicted the existence of two different
classes of quantum critical points for the Kondo problem in graphene,
which was shown to correspond effectively to the problem of a localized
spin $1/2$ coupled to a fermionic bath with electronic density of
states $\rho(\omega)\propto|\omega|^{\bar{r}}$, with $\bar{r}=1$
or $3$. Here we point out that the mean field exponent $\nu=1/3$
derived within the slave boson approach for the class of orbitals
of type II $(\bar{r}=3$) is incorrect. 

Above the upper critical scaling dimension of the Anderson model ($\bar{r}>1$),
the intermediate coupling fixed point is non-interacting and describes
the level crossing between singlet and  doublet states with the
trivial exponent $\nu=1$\cite{Vojta_E}. Albeit fluctuations do not
play a role in the critical behavior for $\bar{r}>1$, the critical
theory is not of the Ginzburg-Landau type and the validity of the
mean-field slave boson equation of state (9) breaks down in the $\bar{r}=3$
class, invalidating Eq. (10), (11), and the inset of Fig. 2b for the
case of type II orbitals. 

All the other results of the paper remain valid, including the spin
exchange Hamiltonian in Eq. (6) and the prediction of a fast power
law for the spatial decay of the RKKY interaction for type II orbitals
($\sim1/R^{7}$). 

We note that since hyperscaling is not obeyed for $\bar{r}>1$, the
scaling prediction for the Kondo temperature with the chemical potential,
$\mu$, in the quantum critical region, $T_{K}\propto|\mu|$, can
be violated\cite{Vojta2_E}. In the situation where the scaling prediction
fails, the criticality in the $\bar{r}=1$ and $\bar{r}=3$ classes
can be in principle distinguished. That will be verified with NRG
methods elsewhere. 

We acknowledge M. Vojta for many helpful discussions.

\end{widetext}


\begin{thebibliography}{10}
\bibitem{note0} In this limit, $J$ is regularized by midgap states
induced by the scattering potential of the impurity. See Ref. \cite{Hentschel}.
\bibitem{note}The RKKY interaction for $H$-sites is not described
by the charge polarization, as suggested in Ref. \cite{Brey}.
\end{thebibliography}
\end{document}